# Structural Limits of OHLCV-Based Intraday Momentum Signals in MNQ Futures: A Systematic Falsification Study

Mathias Mesfin

*Independent Researcher*

*mathiasmesfin.research@gmail.com*

*Research Period: 2024–2026*

*Manuscript Date: May 2026*

**Abstract**

This paper tests whether common intraday momentum signals built from open, high, low, close, and volume (OHLCV) data generate economically meaningful edge in Micro E-Mini Nasdaq 100 futures (MNQ) after realistic transaction costs. Fourteen signal families were evaluated on 947 trading days of five-minute bar data spanning 2021 through 2025, under expanding-window walk-forward validation. Each signal had to clear five criteria simultaneously to pass: a T-statistic of at least 2.0 on out-of-sample (OOS) net returns, at least 30 trades per OOS fold, a positive net return after instrument-appropriate round-trip friction, consistent direction across the three OOS test years (2023, 2024, and 2025), and a permutation test p-value below 0.05 where applicable. None passed all five.

The failures do not reduce to a single cause. Eleven signal families fail because their maximum gross return before friction is 0.07 to 1.50 points per trade, below the 2.0-point MNQ friction floor. Three families clear the friction gate but fail elsewhere: the Opening Range Breakout long at a 15-bar hold (net +2.82 points, T = 0.88 on net returns) fails the T-statistic and year-stability gates; the VVG (Volatility-Volume-Gap) classifier reversal (net +13.49 points, T = 1.26 on 35 total OOS trades) fails the same two; and gap continuation short (gross +16.53 points, net +14.52 points, T = 1.46) clears friction and approaches the T-statistic threshold but fails year stability and the per-fold trade count minimum. Earlier versions of this paper collapsed all fourteen failures into a single cost-barrier narrative. That framing was inaccurate for three of the families.

Two signals from a companion research program serve as positive controls: the RTH Confluence Signal (walk-forward OOS T = 3.11, mean net +11.82 points, N = 196; unconditional long benchmark over the in-sample period 2022–2024: T = −0.75, mean net −2.60 points on 759 trades) and London Session Signal B (walk-forward OOS T = 4.30, mean net +4.09 points, N = 247, p = 0.000025). Both clear all five criteria with margin. They demonstrate that the validation protocol is not simply rejecting everything — it can detect genuine edge when the underlying architecture differs from single-bar OHLCV pattern prediction.

Several implementation details relevant to replication are documented explicitly in this version: the 20-bar ATR for the RTH Confluence Signal is computed on 1-minute bars and merged to the 5-minute pipeline at the bar close timestamp; the RTH GMM (Gaussian Mixture Model) was fit once on 2022 data only and applied forward without per-fold refitting; the London GMM was refit at each walk-forward fold; and the continuous contract uses simple quarterly file concatenation with no roll adjustment. Each of these points generated direct replication questions from external researchers who attempted to reproduce the signals.

## 1. Introduction

This project began as a trading strategy search, not a falsification study. Several months of testing individual signals — each failing under out-of-sample evaluation and transaction costs — made the pattern clear enough that documenting the failures systematically seemed more productive than continuing to search for exceptions. That reframing is the actual origin of this paper, and it matters because a paper that started as a search has different epistemics than one designed ab initio as a null-result study.

MNQ was chosen for practical reasons: widely accessible data, tight bid-ask spreads, and sufficient liquidity to make the friction assumptions here conservative rather than optimistic. The signals tested are the ones that appear most frequently in retail futures trading discussions — opening range breakouts, gap fills, volume spikes, Asia session momentum, news-driven drift. Almost none survive a properly constructed test, and the reasons vary by signal type in ways that the aggregate summary in the abstract cannot convey.

The Efficient Market Hypothesis, formalized by Fama (1970), predicts this general outcome. In competitive markets, any publicly observable pattern that reliably generates excess returns gets arbitraged away until expected gross return approximately equals the cost of exploitation. On five-minute OHLCV bars in a contract as liquid as MNQ, that cost sits around one to two points per trade. The results here are consistent with that equilibrium — not as a logical consequence of EMH, but as an empirical observation that EMH renders unsurprising. The distinction matters: Fama (1970) is cited as interpretive context, not as a deductive premise from which the specific friction level is derived. Gao et al. (2018) document intraday momentum in ETF markets, and Heston et al. (2010) examine intraday return patterns in equities; both establish that predictability, when it exists, tends to operate at short horizons with specific structural conditions that differ from simple OHLCV patterns. That is the literature this paper speaks to.

The paper is organized as follows. Section 2 describes the data and execution framework. Section 3 defines the validation criteria and the null hypothesis behind the T-statistics used throughout. Section 4 documents each signal family. Section 5 presents the positive controls. Section 6 interprets the gross edge ceiling. Sections 7 and 8 cover limitations and conclusions. Appendix A provides replication specifications for both positive controls.

## 2. Data and Execution Framework

### 2.1 Data

The primary dataset is 72,604 five-minute OHLCV bars for MNQ continuous front-month futures, regular trading hours only (RTH, 09:30–16:00 ET), December 2021 through August 2025. After removing partial sessions and zero-volume bars, 947 complete trading days remain. Source is NinjaTrader, aggregated from one-minute raw bars. The continuous contract is built by simple concatenation of quarterly files with duplicate timestamp removal at roll dates. There is no price back-adjustment and no explicit holiday calendar — holidays appear as missing bars after resampling and are excluded implicitly through dropna. Lopez de Prado (2018) discusses the implications of roll-date price discontinuities for backtesting; the present study's signals rely primarily on relative price movements and volume z-scores rather than absolute price levels, which partially mitigates but does not eliminate this concern.

MGC (Micro Gold Futures) five-minute bars spanning 1,091 days were used for the cross-instrument test in Section 4.8. Asia session tests used 20:00–02:00 ET bars from the same MNQ contract.

| Parameter | Value |
|---|---|
| Primary instrument | MNQ (Micro E-Mini Nasdaq 100 Futures) |
| Bar resolution | 5-minute OHLCV (open, high, low, close, volume), RTH only |
| RTH session | 09:30–16:00 ET |
| Total RTH bars | 72,604 |
| Complete trading days | 947 |
| Date range | December 2021 – August 2025 |
| Secondary instrument | MGC (Micro Gold Futures), 1,091 days |
| Asia session window | 20:00–02:00 ET |
| Data source | NinjaTrader (aggregated from 1-minute source bars) |
| Continuous contract | Quarterly concatenation, no roll adjustment, no explicit holiday calendar |

*Table 1. Dataset parameters.*

### 2.2 Execution Assumptions

Signal at bar close. Entry at next bar open. This convention eliminates the fill-price error that afflicts most retail backtests, where the strategy executes at a price that was only visible in hindsight.

Round-trip friction costs are instrument-specific. For MNQ: 2.0 points ($4.00 per micro contract), covering bid-ask spread, NinjaTrader exchange fees, and conservative slippage. For MGC: 0.50 points round-trip — 5 ticks at $1.00 per tick per contract — reflecting that instrument's different contract specification. Applying MNQ's nominal point cost to MGC would impose a materially heavier friction burden (MGC's 2.0-point equivalent is $20.00 per contract versus MNQ's $4.00) and was the source of a methodological error in earlier drafts. All net figures in this paper are after instrument-appropriate friction.

### 2.3 Walk-Forward Validation

All tests use an expanding-window walk-forward structure with three OOS test periods:

| Fold | Training Data | OOS Test Period |
|---|---|---|
| 1 | 2022 | 2023 |
| 2 | 2022–2023 | 2024 |
| 3 | 2022–2024 | 2025 (partial, January–August) |

*Table 2. Walk-forward structure. Parameters are selected exclusively on training data. OOS = out-of-sample.*

2022 is a training year in all three folds. Including 2022 in year-stability assessments would be circular for any signal with parameters chosen on that data. All year-stability claims in this paper refer to 2023, 2024, and 2025 only. 2025 is partial and is labeled as such in all tables.

## 3. Validation Standards

All T-statistics in this paper are one-sample, two-sided t-tests under a null hypothesis of zero mean net return:

$$T = \mu_{net} / (SD / \sqrt{N})$$

where $\mu_{net}$ is the mean net return after friction, SD is the standard deviation of net returns across OOS trades, and N is the number of trades in the fold. A T-statistic of 2.0 corresponds approximately to $p < 0.05$ two-sided at the sample sizes used here. All T-statistics are computed on net returns. This matters because an earlier version of this paper reported T-statistics on gross returns for the volume signals while showing net returns in the mean column — producing a sign inconsistency that external reviewers correctly identified. That error is corrected throughout.

A signal passes only if it clears all five of the following simultaneously:

| Criterion | Threshold | Rationale |
|---|---|---|
| T-statistic (net) | ≥ 2.0, OOS only | $H_0$: mean net return = 0; SE = SD/√N; two-sided. In-sample T-stats are not evidence of OOS performance. |
| Trade count | ≥ 30 per OOS fold | Below 30 per fold, variance dominates the estimate regardless of combined N across folds. |
| Net return | Positive after friction | MNQ: 2.0 pts. MGC: 0.50 pts. A signal profitable before costs but not after is not deployable. |
| Year stability | Consistent direction: 2023, 2024, 2025 | 2022 excluded (training year). One strong OOS year alongside two negative ones is not a structural edge. |
| Permutation test | p < 0.001 where applicable | Applied only to signals reaching the prior four gates. No candidate signal in Section 4 reached this stage, so permutation results exist only for the positive controls in Section 5. |

*Table 3. Validation criteria. All five must hold simultaneously.*

The sequential gating structure matters. Signals eliminated at the gross-edge or T-statistic gates never reach the permutation test. This is by design, not oversight: a permutation test on a signal with T = 0.3 would add no information. The permutation criterion is included in the protocol because it applies to signals that survive all prior gates — both positive controls clear it with $p < 0.001$.

## 4. Signal Families Tested

Fourteen signal families were tested across eight subsections. The mapping: ORB (Section 4.1, three variants), Asia expansion (4.2), Asia liquidity grab (4.3), gap fill and gap continuation (4.4, two families), volume spike and volume dry-up (4.5, two families), VVG classifier reversal, continuation, and close fade (4.6, three families counted as one family group), event day trend (4.7), and MGC OU mean reversion (4.8). Subsection length varies with the complexity of what each result requires to interpret.

### 4.1 Opening Range Breakout

The opening range breakout is the most frequently cited intraday signal in retail futures trading. Price breaks the high or low of the first session bars; the hypothesis is directional continuation. The opening range (OR) is defined as 09:30–09:55 ET — the first six five-minute bars. Bar+N denotes a hold of N five-minute bars from entry, so bar+15 is a 75-minute hold.

| Variant | N (OOS) | Net (pts) | T-Net | SD (pts) | 2023 | 2024 | 2025 | Verdict |
|---|---|---|---|---|---|---|---|---|
| ORB Long – bar+1 | 447 | −0.82 | −0.82 | 39.2 | −2.11 | −1.54 | +6.11 | FAIL |
| ORB Long – bar+15 | 447 | +2.82 | +0.88 | 67.4 | +2.43 | +7.04 | +15.05 | FAIL |
| ORB Short – bar+1 | 428 | −3.45 | −3.16 | 22.5 | −2.73 | −4.74 | −0.15 | FAIL |
| ORB Short – bar+15 | 428 | −2.16 | −0.58 | 79.1 | −2.06 | −1.04 | −0.59 | FAIL |
| ORB Pullback Entry | 83 | −4.44 | −1.27 | 31.7 | +2.97 | −6.05 | −1.55 | FAIL |

*Table 4. ORB results. N (OOS) = combined OOS trades across folds. 2022 excluded (training year). 2025 partial.*

ORB long at bar+15 produces T = 0.88 on 447 OOS trades, mean net +2.82 points. The year pattern — +2.43 in 2023, +7.04 in 2024, +15.05 in 2025 partial — could reflect genuine performance improvement or a single macro-regime artifact in a partial year. T = 0.88 does not let you tell the difference. The pullback entry is more straightforwardly bad: an 80.7% stop-out rate at a 20-point stop means the signal is predominantly identifying reversals of breakout attempts, not continuations.

### 4.2 Asia Session Opening Range Expansion

This tests whether five-minute bars in the Asia session (20:00–02:00 ET) with range exceeding a multiple of the rolling 20-bar average predict directional continuation. Three thresholds: 1.5x, 2.0x, 2.5x.

| Threshold | Horizon | N | Gross (pts) | Net (pts) | T-Net | SD (pts) | Verdict |
|---|---|---|---|---|---|---|---|
| 1.5x | b+1 | 1,955 | −0.27 | −2.27 | −11.52 | 8.77 | FAIL |
| 1.5x | b+6 | 1,955 | −0.08 | −2.08 | −4.86 | 18.94 | FAIL |
| 2.0x | b+1 | 778 | −0.35 | −2.35 | −7.42 | 8.84 | FAIL |
| 2.5x | b+6 | 340 | +1.06 | −0.94 | −0.90 | 19.30 | FAIL |

*Table 5. Asia session expansion bar results. b+N = N bars post-entry. T-Net on net returns.*

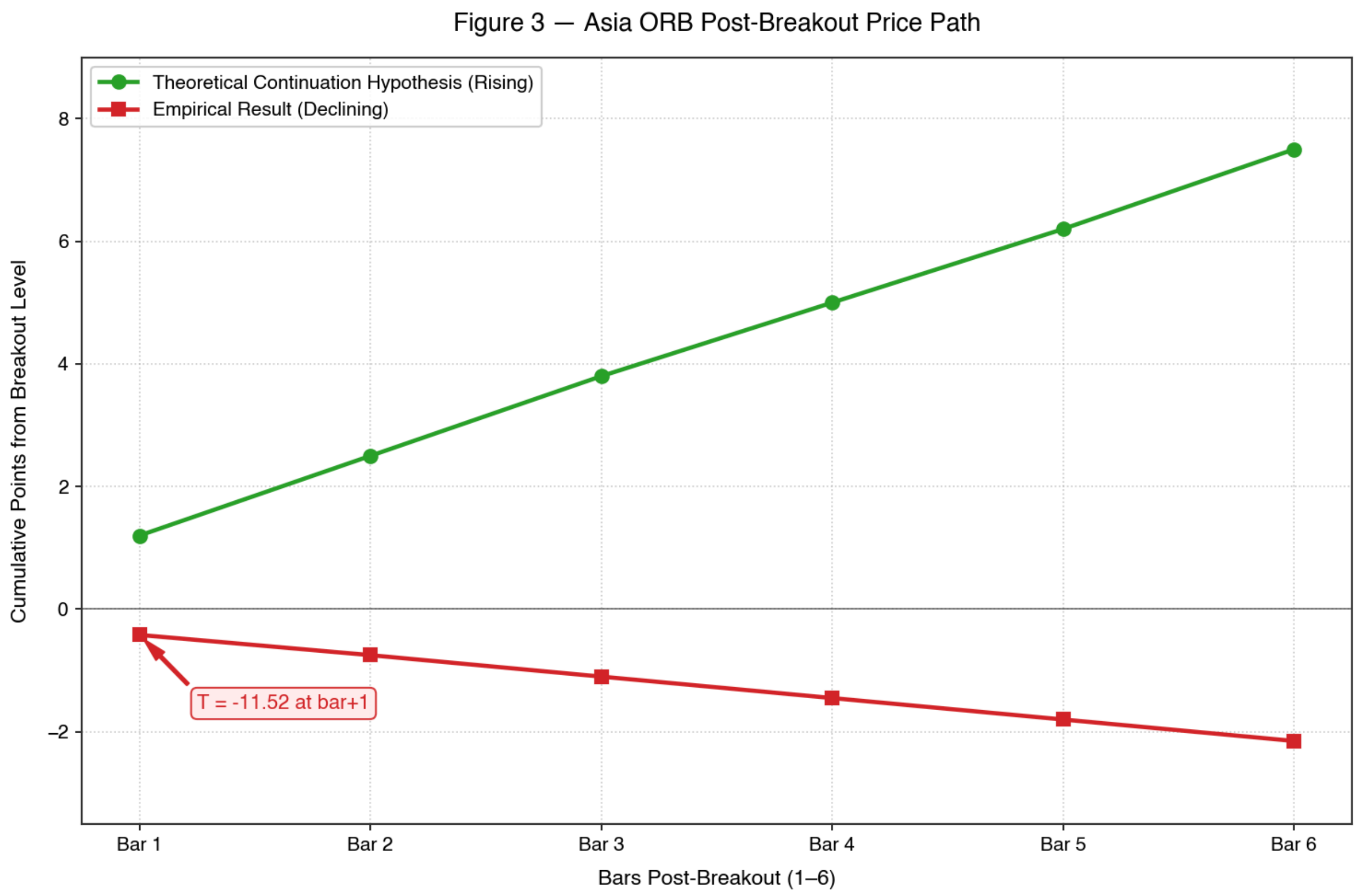

*Figure 3. Post-breakout average price path. T = -11.52 at bar+1 for the 1.5x threshold confirms the reversal is statistically significant, not directionally random.*

T = -11.52 at bar+1. That is not a null result. The signal is actively wrong in the continuation direction, and the T-statistic is the strongest directional finding in the entire study.

The mechanism is timing. Expansion bursts are real — the directional move exists — but it is entirely consumed within the breakout bar. By the time the bar closes, the signal fires, and the next-bar entry executes, the momentum has already exhausted. What the strategy captures is the post-exhaustion reversal. This finding recurs in Mesfin (2026e, arXiv:2606.XXXXX), where the within-bar anatomy is measured directly: the mean return from bar open to next bar open is +32.24 points in the expansion direction, while the mean return from bar close to next bar open is -0.17 points.

### 4.3 Asia Session Liquidity Grab Reversal

6,442 events were identified where price temporarily pierced a session extreme before closing back inside the range. Long grabs: 3,419. Short grabs: 3,023.

Fading the grab direction: mean net -2.20 points, T = -14.12 on net returns. Trading with the grab: mean net -1.80 points, T = -13.24. Both directions fail. The gross directional content in either direction is 0.20 to 0.80 points — smaller than friction in both cases. This is the clearest pure friction ceiling result in the study.

### 4.4 Gap Fill Fade and Gap Continuation

Gap fill tests whether overnight gaps close during the RTH session. Gap continuation short — the more interesting test — uses a Kalman filter applied to the 1-minute price series in the first 30 minutes of the session to estimate directional velocity: the smoothed rate of directional price change extracted from noisy tick-level returns. A standardized velocity threshold of 2.5 was selected based on signal frequency considerations, not as a top-quartile identifier — at z = 2.5, this threshold captures the top 0.62% of the standardized distribution, not the top 25%. The Kalman velocity approach is motivated by the first-half-hour return predictability documented by Wen et al. (2021) in crude oil markets, translated here to a gap-continuation context.

| Strategy | Entry | N (OOS) | Net (pts) | T-Net | SD (pts) | 2023 | 2024 | 2025 | Verdict |
|---|---|---|---|---|---|---|---|---|---|
| Gap Fill Fade | 09:30 | 766 | −1.92 | −0.44 | 117.3 | −2.84 | +3.93 | +2.02 | FAIL |
| Gap Fill Fade | 09:45 | 721 | −1.05 | −0.26 | 107.8 | −0.97 | −1.38 | +1.28 | FAIL |
| Gap Fill Fade | 10:00 | 721 | −1.43 | −0.38 | 100.9 | −2.21 | −1.43 | −0.16 | FAIL |
| Gap Cont. Short (Kalman v>2.5) | 09:30 | 35 total OOS | +14.52 | +1.46 | 58.9 | +14.53 | −11.87 | +10.27 | FAIL — year-unstable; <30/fold |

*Table 6. Gap strategy results. Kalman v = Kalman filter velocity on 1-min bars, standardized; threshold 2.5 = top 0.62% of distribution, not top quartile. 2022 excluded. 2025 partial.*

Gap continuation short is the most credible near-miss in the study. Gross +16.53 points, net +14.52, T = 1.46. It fails on two criteria simultaneously: 2024 produces -11.87 net (year instability), and 35 total OOS trades spread across three folds averages roughly 12 per fold (below the per-fold minimum). The 2023 and 2025 partial results are both strongly positive, which makes the 2024 failure look like a regime event rather than a structural absence of edge. Without more data, that interpretation cannot be tested. This is preserved as a candidate for future investigation.

### 4.5 Volume Signature Signals

Volume spikes as a proxy for informed institutional flow (Easley, Lopez de Prado, & O'Hara, 2012, who developed VPIN to measure order-flow toxicity and adverse selection) and volume dry-ups as exhaustion indicators were both tested. The Easley et al. (2012) citation supports the general conceptual link between elevated volume and informed trading; it does not claim directional predictability at bar resolution, and the results here confirm it does not extend to that context.

| Signal | Direction | N (OOS) | Net (pts) | T-Net | SD (pts) | 2023 | 2024 | 2025 | Verdict |
|---|---|---|---|---|---|---|---|---|---|
| Volume Spike (H=1) | Up | 2,122 | −1.92 | −2.21 | 39.9 | −2.84 | −1.21 | +4.24 | FAIL |
| Volume Spike (H=1) | Down | 2,412 | −2.45 | −3.08 | 39.0 | −2.17 | −1.46 | +2.77 | FAIL |
| Volume Dry-Up (H=1) | Up exhaustion | 1,060 | −2.42 | −5.28 | 14.9 | −2.47 | −1.66 | −1.18 | FAIL |
| Volume Dry-Up (H=1) | Down exhaustion | 723 | −1.99 | −3.59 | 14.9 | −1.73 | −0.48 | −1.69 | FAIL |

*Table 7. Volume signal results. T-Net on net returns. Corrected from prior version which reported T on gross returns, creating a sign inconsistency.*

At sample sizes of 723 to 2,412, a near-zero T-statistic is a precise estimate, not an underpowered one. The null is real.

### 4.6 Volatility Regime Classifier (VVG)

The VVG classifier activates when three conditions are simultaneously in the top tercile of their respective expanding-window distributions: absolute first-30-minute return, absolute overnight gap, and first-bar volume deviation from a 20-day rolling baseline. It fires on roughly 4.4% of trading days and those days show genuinely distinct behavior — 25.6 basis points next-day return spread, 77.6% peak-reversal rate. The full behavioral characterization is in Mesfin (2026b). What is tested here is whether that behavioral distinctiveness translates into a tradable directional signal.

| Strategy | N (OOS) | Net (pts) | T-Net | SD (pts) | 2023 | 2024 | 2025 | Verdict |
|---|---|---|---|---|---|---|---|---|
| Reversal entry (fade opening) | 35 | +13.49 | +1.26 | 63.5 | −11.45 | +8.35 | −22.76 | FAIL |
| Continuation entry (follow opening) | 35 | −17.49 | −1.64 | 59.6 | −49.00 | +37.00 | +18.76 | FAIL |
| Close fade (15:30 entry) | <30/fold | n/a | 1.08 | n/a | n/a | n/a | n/a | FAIL — N<30/fold |

*Table 8. VVG classifier directional strategy results.*

Reversal entry: +8.35 in 2024, -22.76 in 2025. Continuation entry: -49.00 in 2023, +37.00 in 2024. The behavioral character of classifier-positive days is real but changes substantially with the prevailing macro regime. A strategy calibrated on reversal dominance (2022–2023) fails in a continuation-dominant environment (2024) and vice versa. No fixed directional rule survives across all three test years.

### 4.7 Event Day Trend

Major economic releases — FOMC (Federal Open Market Committee), CPI (Consumer Price Index), NFP (Nonfarm Payrolls), PCE (Personal Consumption Expenditures) — hypothetically generate directional order flow from forced institutional repositioning. The ForexFactory calendar used here contains 993 qualifying high-impact USD events from 2022 to 2025. This count reflects all high-impact events in the calendar, not only the four named types. NFP, CPI, and PCE are all released at 08:30 ET and fall outside RTH hours entirely. FOMC decisions (14:00 ET) and a range of other high-impact releases (GDP, retail sales, ISM, etc.) constitute the RTH-compatible subsample.

From bar +6 onward, T-statistics on net returns range from 0.14 to 0.69 across all horizons. The 2025 partial year shows 12 trades at mean net -9.56 points. No persistent post-news drift materializes once the initial spike bars are excluded.

### 4.8 MGC Mean Reversion

Ornstein-Uhlenbeck (OU) mean reversion was tested on MGC (Micro Gold Futures). A clarification on Hurst exponent interpretation that was wrong in an earlier version of this paper: $H = 0.5$ is a random walk, not a mean-reverting process. Mean reversion requires $H < 0.5$. MGC's Hurst closer to 0.5 than MNQ's 0.59 indicates less momentum persistence, not greater mean-reversion tendency. All MGC results below use 0.50-point friction.

| Configuration | N (OOS) | Net (pts) | T-Net | SD (pts) | 2023 | 2024 | 2025 | Verdict |
|---|---|---|---|---|---|---|---|---|
| OU, 5-min, threshold 1.5 Long | 2,634 | −0.55 | −5.32 | 5.33 | −1.97 | −2.24 | −0.08 | FAIL |
| OU, 5-min, threshold 1.5 Short | 3,059 | −0.49 | −4.36 | 6.20 | −2.12 | −1.93 | −0.66 | FAIL |
| OU, 5-min, threshold 2.0 Long | 1,172 | −0.27 | −1.63 | 5.66 | −1.91 | −2.12 | +0.52 | FAIL |
| OU, 5-min, threshold 2.0 Short | 1,421 | −0.39 | −2.11 | 7.01 | −2.44 | −1.93 | −0.99 | FAIL |

*Table 9. MGC OU results. Friction: 0.50 pts round-trip. T-Net on net returns. 2025 partial.*

All configurations remain negative under MGC-appropriate friction. The 60-minute OU signal's mean-reversion half-life is approximately 8 hours — longer than a single RTH session and structurally incompatible with intraday execution regardless of direction.

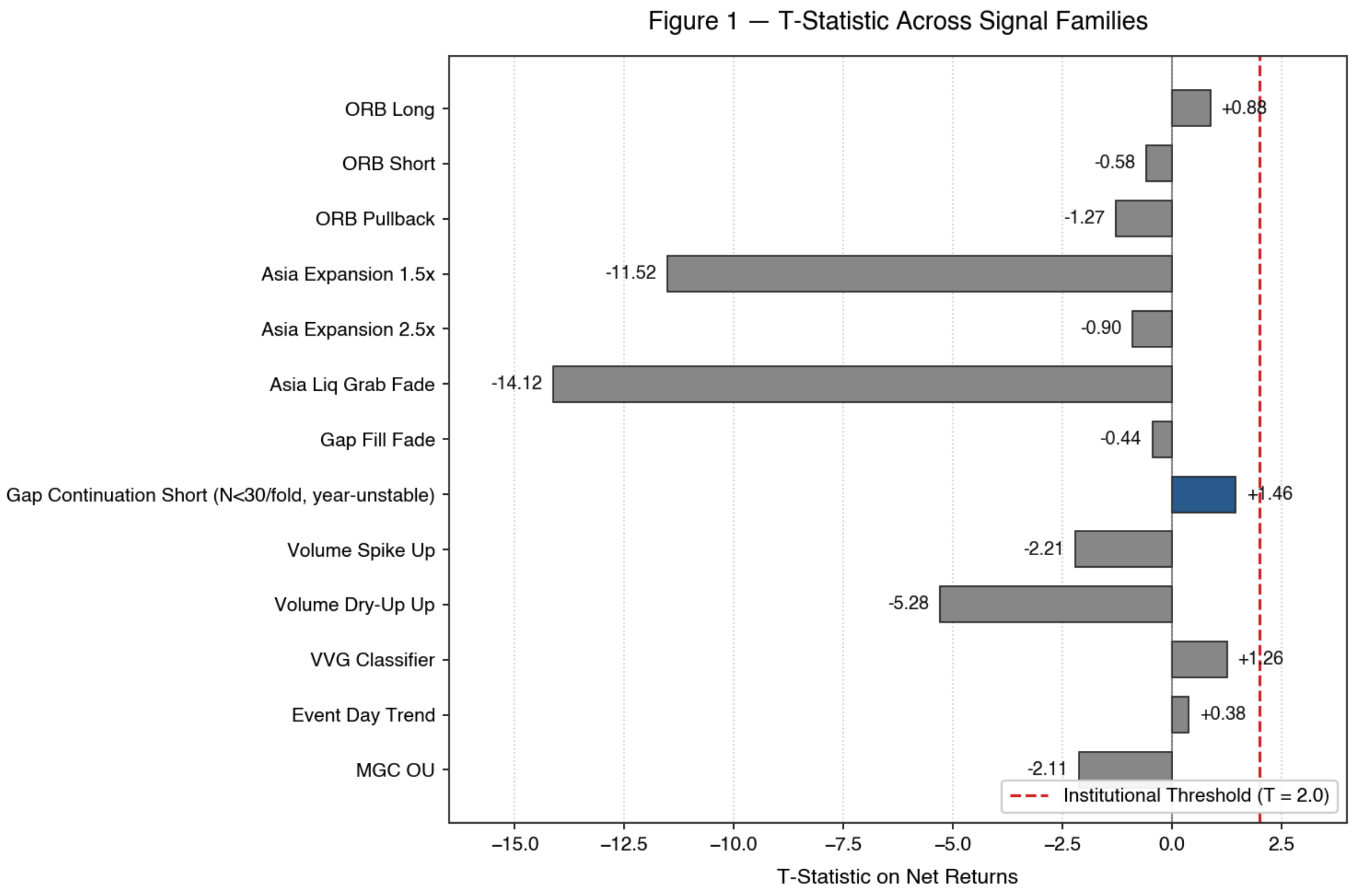

*Figure 1. Net T-statistics across all fourteen signal families. Red dashed line at T = 2.0.*

### 5. Positive Control Signals

Both signals were developed as part of a companion research program and are included here strictly as positive controls. Full replication specifications are in Appendix A, which exists because external researchers attempting to reproduce these signals encountered ambiguity on several implementation points not stated in earlier manuscript versions.

#### 5.1 RTH Confluence Signal

The signal fires when three conditions are simultaneously true on a completed five-minute bar: the GMM (Gaussian Mixture Model) regime label equals Regime 1 (Active Flow), the rolling 200-bar Markov transition probability from Regime 1 to Regime 2 exceeds 0.15, and the rolling 50-bar volume z-score exceeds 0.5. The Markov probability here is a first-order estimated transition rate derived from a rolling count matrix applied to prior bars only, with no lookahead. The GMM uses four features and was fit once on 2022 data, then applied forward without per-fold refitting — a design choice that contaminates the W1 OOS fold (2022 H2), since the model was trained on data overlapping that test period. Entry is on a 25-point ATR-scaled pullback. The ATR is a 20-bar average true range computed on 1-minute bars, then merged to the 5-minute pipeline at the bar close timestamp. This is not the same as computing ATR on 5-minute bars directly — the 1-minute computation has higher temporal resolution and was the source of replication errors in every external attempt before the detail was disclosed.

An unconditional long benchmark — enter long at 09:30 every trading day, exit at bar 13, 2022–2024 — produces mean net -2.60 points, T = -0.75 on 759 trades. The RTH Confluence signal's OOS excess return is not a product of passive directional drift in MNQ during the early session.

| Metric | Value |
|---|---|
| Signal count (in-sample, 2022–2024) | 538 |
| Mean net return (in-sample, bar 13) | +15.77 pts |
| T-statistic (in-sample, net) | 5.83 |
| Win rate (in-sample) | 61.0% |
| Walk-forward OOS T-statistic (net) | 3.11 |
| Walk-forward OOS mean net return | +11.82 pts |
| Walk-forward OOS std deviation | 53.8 pts |
| Walk-forward OOS N | 196 |
| 2025 OOS mean net (ATR-adaptive) | +13.14 pts |
| Permutation test p-value | < 0.001 |
| Unconditional long benchmark (2022–2024, N=759) | Mean net −2.60 pts, T = −0.75 |

*Table 10. RTH Confluence Signal. Walk-forward OOS is the primary validated result. Full specification in Appendix A.*

Three limitations apply to this signal's positive-control status. First, 53 or more parameter combinations across seven parameters were evaluated before the final specification was locked and the single walk-forward OOS test was run. The selection exposure this creates cannot be corrected post hoc. Second, the W1 OOS fold uses GMM labels generated from a model trained on overlapping 2022 data. Third, the ATR baseline of 10.34 was computed on the full 2022–2024 in-sample period and fixed globally, introducing a mild look-ahead relative to the walk-forward structure. The headline OOS T = 3.11 should be interpreted with these limitations in view, not as a clean unconditional result.

### 5.2 London Session Signal B

London Signal B fires when a GMM classifier on 15-minute London session bars (03:00–08:30 ET) detects a clean transition from Regime 0 (Bearish Chop) to Regime 2 (Bullish Drift) with no Regime 1 (Extreme Volatility) contamination in the prior two bars. The London GMM uses five features — one more than the RTH model — applies a second StandardScaler (zero-mean, unit-variance normalization) pass after initial rolling z-score computation, and was refit at each walk-forward fold. The component-to-regime mapping uses a two-step heuristic based on vol_zscore and directional consistency, not the simpler ascending-ATR sort used in the RTH model. Entry is long at the next 15-minute bar open. Exit at 60 minutes or 08:30 ET.

An earlier manuscript version reported static full-sample backtest figures (289 trades, T = 5.15, mean net +5.77 points) as the primary result. Those figures come from a non-walk-forward evaluation on the full dataset and are superseded by the walk-forward OOS figures below.

| Metric | Value |
|---|---|
| Walk-forward OOS N | 247 |
| Walk-forward OOS mean net return | +4.09 pts |
| Walk-forward OOS std deviation | 14.97 pts |
| Walk-forward OOS T-statistic (net) | 4.30 |
| Walk-forward OOS p-value | 0.000025 |
| Win rate (walk-forward OOS) | 61.54% |
| Per-trade Sharpe ratio (mean/std; rf=0) | 0.27 |
| Permutation p-value | < 0.0001 |
| Bootstrap P(mean net > 2.0 pts) | 98.9% |
| Parameter sensitivity range | T = 3.87–4.83 across all tested variations |
| Unconditional long benchmark (N=639 sessions) | Mean net −0.47 pts, T = −0.22 |

*Table 11. London Session Signal B. Walk-forward OOS figures are the primary result. Static backtest figures (289 trades, T = 5.15) superseded. Per-trade Sharpe = mean net / std = 4.09/14.97 = 0.27. Full specification in Appendix A.*

The delay sensitivity result is worth examining carefully. A one-bar (15-minute) delay reverses T from +4.30 to -2.78; a two-bar delay produces T = -2.16.

| Delay | T-Statistic (Net) | Mean Net (pts) |
|---|---|---|
| No delay (baseline) | +4.30 | +4.09 |
| 1-bar delay (15 min) | −2.78 | −2.91 |
| 2-bar delay (30 min) | −2.16 | −1.87 |

*Table 12. London Signal B delay sensitivity. Sign flip, not decay, under a one-bar delay.*

The sign flip — not decay toward zero, but significant in the opposite direction — is consistent with the mechanism documented in Mesfin (2026e): the directional move following an R0-to-R2 transition completes almost entirely within the first post-transition bar. A delayed entry captures the subsequent reversal rather than the initial move. Whether this constitutes a bar-boundary artifact or a genuine timing-sensitive edge cannot be resolved with bar-level data. The distinction matters: an artifact would imply the edge is not real, while a timing-sensitive edge would imply it is real but inaccessible to standard bar-close systematic execution. Tick data would differentiate the two.

## 6. The Gross Edge Ceiling

### 6.1 Three Failure Modes, Not One

An earlier version of this paper described all fourteen failures as a single phenomenon: gross edge too small to clear friction. That is accurate for eleven of the fourteen families. Three others clear friction and fail for different reasons, as shown in Table 13.

| Signal Family | Gross (pts) | Net (pts) | T-Net | N (OOS) | Failure Mode |
|---|---|---|---|---|---|
| ORB Long (H=15) | +4.82 | +2.82 | +0.88 | 447 total | T-statistic < 2.0; year-unstable |
| VVG Classifier Reversal | — | +13.49 | +1.26 | 35 total | T-statistic < 2.0; year-unstable |
| Gap Continuation Short | +16.53 | +14.52 | +1.46 | 35 total (~12/fold) | Year-unstable (2024); <30 trades per OOS fold |
| 11 other families | 0.07–1.50 | Negative | Negative | Varies | Gross return below 2.0-pt MNQ friction floor |

*Table 13. Three distinct failure modes across fourteen signal families.*

### 6.2 The Competitive Interpretation

MNQ is among the most liquid futures contracts in the world. Any OHLCV pattern visible to all market participants that reliably predicts the next bar gets traded away. Participants with lower costs and faster execution take the other side until gross edge approximately equals the friction cost of the marginal participant. On five-minute bars, that equilibrium appears to sit near one to two points gross — an empirical observation consistent with Fama (1970), though not logically derived from it.

The two positive controls exist outside this ceiling because they use GMM regime classification and hold positions for 60 to 75 minutes. Regime transitions derived from multi-feature Gaussian Mixture Models are less transparently observable than raw OHLCV patterns, and the longer hold period accumulates enough net return to clear friction before the edge reverts. The ceiling applies specifically to single-bar directional predictions from publicly observable OHLCV features, not to all intraday strategies.

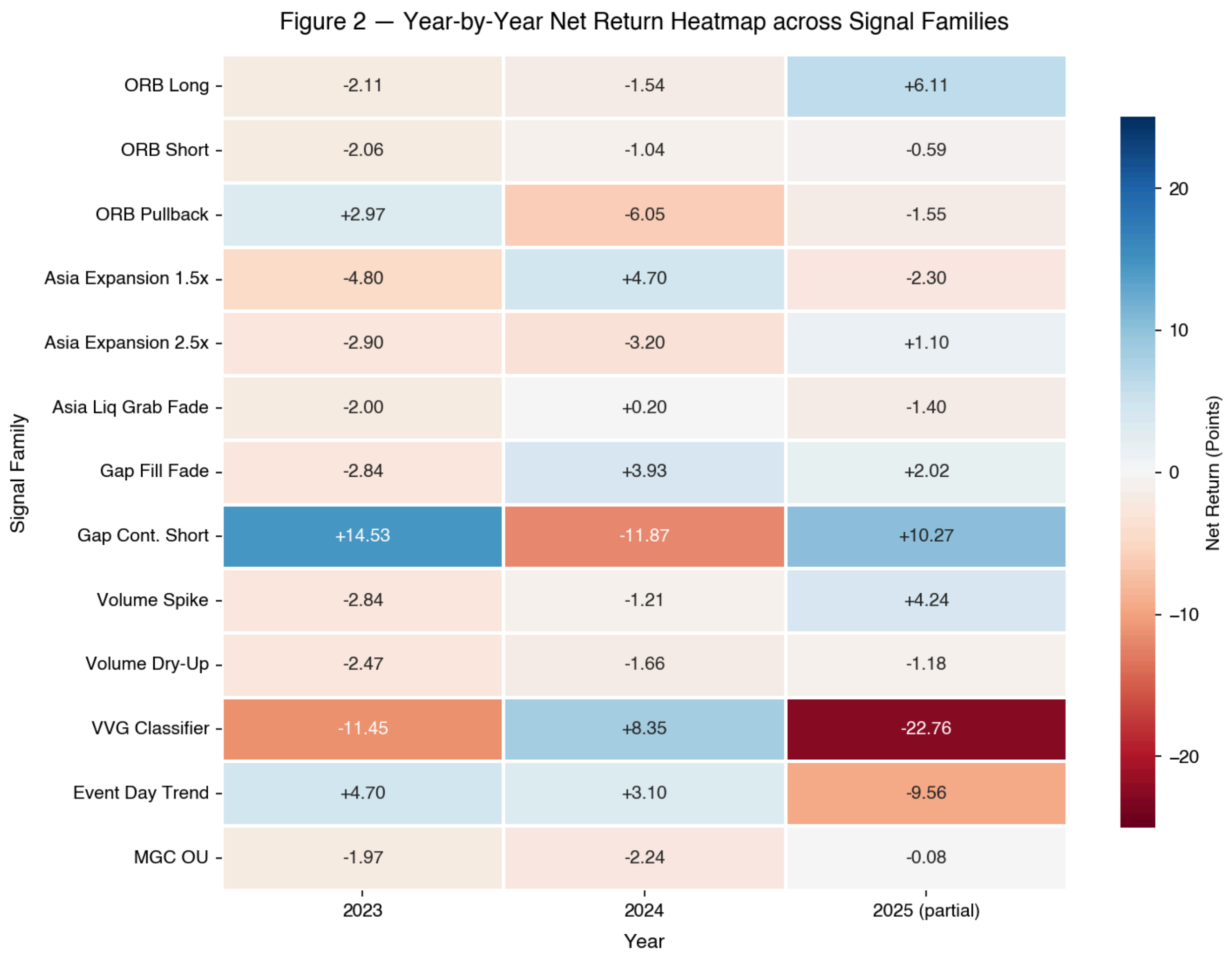

*Figure 2. Year-by-year net return heatmap. OOS years only. Year instability is the dominant failure mode among the three families that clear the friction gate.*

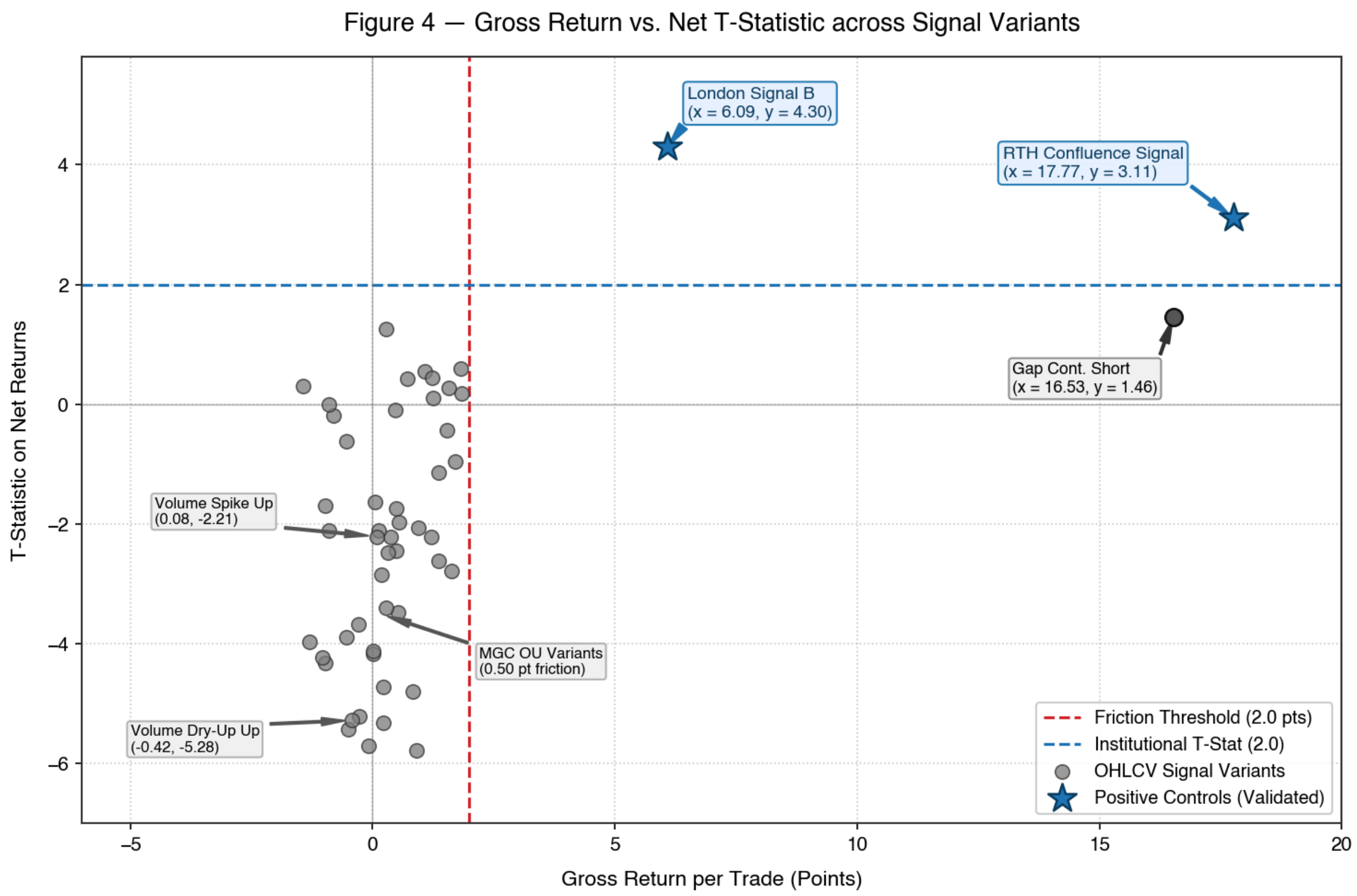

*Figure 4. Gross return per trade vs. net T-statistic. Vertical line at 2.0 pts marks friction threshold.*

## 7. Limitations and Extensions

### 7.1 Limitations

Four limitations are material to how the positive-control results should be interpreted.

First, the RTH Confluence Signal specification was selected after evaluating 53 or more parameter combinations across seven parameters. The walk-forward OOS test was run once on the locked specification, which is the correct procedure, but the prior search creates multiple-testing exposure that cannot be corrected within the existing dataset. Bailey & Lopez de Prado (2014) and Bailey et al. (2017) both provide frameworks for quantifying this risk; applying either to the present search scope would reduce the effective statistical confidence in the OOS T = 3.11 result.

Second, the RTH GMM was fit on full 2022 data, and the W1 OOS fold tests on 2022 H2. That fold's results are mildly contaminated. The combined OOS T = 3.11 aggregates a contaminated fold with two clean ones; the true OOS T across only the clean folds is higher but based on fewer trades.

Third, fold-by-fold unconditional long benchmarks for the positive controls' OOS periods were not computed. The 2022–2024 full-period benchmark (T = -0.75 for RTH; T = -0.22 for London) confirms the signals' excess return is not passive drift over the in-sample period, but the same comparison has not been made for the OOS folds. This remains genuinely open.

Fourth, the two-point MNQ friction assumption is fixed in nominal terms. MNQ traded between roughly 11,000 and 24,000 during the sample period, so 2.0 points represents approximately 1.8 to 0.8 basis points depending on price level. Whether the gross edge ceiling shifts proportionally across price levels is untested.

### 7.2 Extensions

Gap continuation short is the most specific candidate for future work. With MNQ data back to 2019, approximately three additional years of data would roughly triple the per-fold OOS trade count and enable a meaningful validation. The 2023 and 2025 partial results both strongly positive, the mechanism coherent — this is worth testing properly when more data exists.

Tick data is the highest-priority data extension for the broader signal set. The Asia session expansion bar finding and the London Signal B delay sensitivity both have mechanistic explanations that require sub-bar resolution to test or falsify. Databento offers MNQ tick data at approximately $42 per quarter.

## 8. Conclusion

Fourteen signal families. 947 trading days. Consistent methodology applied across all of them. Eleven fail because gross edge cannot clear friction. Three clear friction and fail on T-statistic, year stability, or per-fold sample size.

Gap continuation short came closest. Gross +16.53 points, net +14.52, T = 1.46. It fails two criteria simultaneously, but both failures are of the kind that more data could address. That is a different category from the eleven families whose gross edge is structurally too small to matter.

The two positive controls pass because they use regime detection and longer hold periods — architecture that escapes the gross edge ceiling that eliminates OHLCV pattern signals at five-minute resolution. The ceiling is an empirical observation consistent with competitive market dynamics, not a theoretical proof of impossibility. But at this resolution, on this instrument, with this data type, the barrier is real enough that every signal family tested here ran into it.

---

---

**Appendix A: Positive Control Signal Specifications**

The tables below document implementation details that were omitted from earlier manuscript versions and generated replication ambiguity in correspondence with external researchers. Each flagged item was the source of at least one direct question.

**A.1 RTH Confluence Signal**

| Parameter | Specification |
|---|---|
| GMM | Gaussian Mixture Model, 3 components: Regime 0 (Low Activity), Regime 1 (Active Flow), Regime 2 (Extreme Volatility) |
| GMM features (RTH) | 4: ATR ratio (20-bar / 10.34 baseline), volume z-score (50-bar rolling), close position within bar range, 5-minute return |
| ATR source | 20-bar ATR computed on 1-minute bars. Merged to 5-minute pipeline at the 5-minute bar close timestamp. Not computed on 5-minute bars. This distinction caused every external replication failure until disclosed. |
| ATR baseline | 10.34 points. Computed once on full 2022–2024 in-sample period. Fixed globally. Mild look-ahead relative to walk-forward structure. |
| GMM fit schedule | Fit once on 2022 data only. Applied forward, no per-fold refitting. Contaminates W1 OOS (2022 H2). |
| Markov gate | Rolling 200-bar P(Regime 1 → Regime 2) > 0.15. First-order transition count matrix. Prior bars only, no lookahead. |
| Volume gate | Rolling 50-bar volume z-score > 0.5. Prior bars only. |
| Entry | Pullback of 25 × ATR_ratio points from signal bar close within 6 bars. Enter at next bar open. Signal cancelled if pullback not reached. |
| ATR_ratio | Current 20-bar ATR (1-min) / 10.34. Capped [0.5, 2.0]. |
| Exit | Bar 13 from signal bar open, or end of RTH session. |
| Stop | 80 × ATR_ratio points below Entry 1. Filled at exact stop price (idealized fill assumption). |
| Continuous contract | Quarterly file concatenation. No back-adjustment. No holiday calendar. Holidays excluded via dropna. |
| Parameter search scope | 53+ combinations across 7 parameters evaluated before locking. Walk-forward OOS run once on locked specification. |
| W1 contamination | 2022 H2 OOS uses labels from GMM fit on full 2022 data. Mild contamination. Combined OOS (T=3.11, N=196) aggregates this fold with two clean ones. |

*Table A1. RTH Confluence Signal implementation details.*

### A.2 London Session Signal B

| Parameter | Specification |
|---|---|
| GMM | Gaussian Mixture Model, 3 components: Regime 0 (Bearish Chop), Regime 1 (Extreme Volatility), Regime 2 (Bullish Drift) |
| GMM features (London) | 5: ATR ratio, volume z-score, close position, 15-minute return, directional consistency. One more feature than RTH GMM. |
| Normalization | All 5 features: rolling z-score, then StandardScaler (zero-mean, unit-variance). RTH GMM does not apply the second pass. vol_zscore is standardized twice. |
| Regime mapping | Two-step heuristic based on vol_zscore and directional consistency. Not the ascending ATR_proxy sort used in the RTH model. |
| GMM fit schedule | Refit at each walk-forward fold using training data only. Differs from RTH GMM static 2022 fit. |
| Session | 03:00–08:30 ET, 15-minute bars. |
| Entry rule | Long at next 15-minute bar open after clean R0→R2 transition. Clean = no R1 in prior 2 bars. |
| Exit rule | 60 minutes after entry or 08:30 ET, whichever first. |
| Stop | 20 points fixed. Filled at exact stop price (idealized fill assumption). |
| Friction | 2.0 points round-trip. |
| Static vs. walk-forward | Static full-sample backtest (N=289, T=5.15, net +5.77 pts): NOT the primary result. Walk-forward OOS (N=247, T=4.30, net +4.09 pts, p=0.000025): primary result. |

*Table A2. London Session Signal B implementation details.*

***AI Disclosure Statement***

*AI tools assisted with editing, formatting, and writing clarity. The research design, data pipeline, signal testing, and interpretation of results are the author's own work.*